# Spin-polarized angle-resolved photoelectron spectroscopy of the so-predicted Kondo topological insulator $SmB_6$

Shigemasa Suga[1,2*], Kazuyuki Sakamoto[3], Taichi Okuda[4], Koji Miyamoto[4], Kenta Kuroda[5], Akira Sekiyama[6], Junichi Yamaguchi[6], Hidenori Fujiwara[6], Akinori Irizawa[1], Takahiro Ito[7+], Shinichi Kimura[7#], T. Balashov[8], W. Wulfhekel[8], S. Yeo[9], Fumitoshi Iga[10] and Shin Imada[11]

[1]Institute of Scientific and Industrial Research, Osaka University, 567-0047 Osaka, Japan
[2]Max-Planck-Institute for Microstructure Physics, Weinberg 2, Halle 06120 Germany
[3]Department of Nanomaterials Science , Chiba University, 263-8522 Chiba, Japan
[4]Hiroshima Synchrotron Radiation Center, Hiroshima University, 2-313 Kagamiyama, Higashi-Hiroshima 739-0046, Japan
[5]Graduate School of Science, Hiroshima University, 1-3-1 Kagamiyama, Higashi-Hiroshima 739-8526, Japan
[6]Graduate School of Engineering Science, Osaka University, 1-3 Machikaneyama, Toyonaka, Osaka 560-8531, Japan
[7]UVSOR, Institute for Molecular Science, Okazaki, 444-8585 Aichi, Japan
[8]Physikalisches Institut, Karlsruhe Institute of Technology, 76131 Karlsruhe, Germany
[9]Korea Atomic Energy Research Institute, Daejeon 305-600, Republic of Korea
[10]Faculty of Science, Ibaraki University, Mito 310-0056, Ibaraki, Japan
[11]Department of Physical Sciences, College of Science and Engineering, Ritsumeikan University, 1-1-1 Nojihigashi, Kusatsu, 525-8577 Shiga, Japan

*all correspondence should be addressed to ssmsuga@gmail.com
+Present address Nagoya University Synchrotron Radiation Research Center and Graduate School of Engineering, Nagoya University 464-8603 Aichi, Japan
#Present address: Graduate School of Frontier Biosciences, Osaka University, 1-3 Yamadaoka, Suita, Osaka 565-0871 Japan

Keywords: **Spin-polarized photoelectron, Angle-resolved photoemission, low temperature, Kondo Semiconductor, Topological insulator, bulk, surface**

## Abstract

Undoped and slightly Eu-doped $SmB_6$ show the opening of a gap with decreasing temperature below ~150 K. The spectral shapes near the Fermi level ($E_F$) at 15 K have shown strong increase in intensity of a peak at a binding energy ($E_B$) of around 12 meV with decreasing the photon energy (hν) from 17 eV down to 7 eV. Angle resolved spectra of $SmB_6$ measured at hν=35 eV just after the in-situ cleavage showed clear dispersions of several bands in the $E_B$ region from $E_F$ to 4 eV. Spin-polarized photoelectron spectra were then measured at 12 K and light incidence angle of ~50°. In contrast to the lack of spin polarization for the linearly polarized light excitation, clear spin polarization was observed in the case of circularly polarized light excitation. The two prominent peaks at $E_B$~12 and ~150 meV have shown opposite signs of spin polarization which are reversed when the helicity of the light is reversed. The sign and the magnitude of spin-polarization are consistent with a theoretical prediction for the $^6H_{5/2}$ and $^6H_{7/2}$ states.

## 1. Introduction

$SmB_6$ is known to show a gap opening at low temperatures below ~150 K [1,2], while it is a mixed valence metallic material above this temperature. Although Kondo semiconductor scenario was proposed for this material as well as for $YbB_{12}$ systems at low temperatures [3,4], the different behaviors of the change in the spectral shapes between these two materials were experimentally clarified [5]. Very recently, a possibility of a topological insulator scenario was theoretically proposed for $SmB_6$ [6,7]. In parallel, intensive studies of angle-resolved photoelectron spectroscopy (ARPES) are going on in several groups [8–12]. Efforts are also made for spin-polarized angle-resolved photoelectron spectroscopy (SP-ARPES) of $SmB_6$.

Clarification of the surface and bulk electronic structure is quite important in such a study because the surface electronic structure is often very different from the bulk electronic structure in most lanthanide compounds. The photon energy (hν) dependent study is powerful for this purpose owing to the variation of the probing depth from the sample surface with hν (or kinetic energy $E_K$) and the hν dependence of the photoionization cross section (PICS) of different orbitals. In addition, full utilization of the circularly and linearly polarized light is powerful to discuss the origin of the observed spin polarization of photoelectrons [13,14].

## 2. Experimental

The angle-integrated photoelectron spectra of $Sm_{0.85}Eu_{0.15}B_6$ and $SmB_6$ were measured on fractured surfaces at temperatures between 200 and 15 K at BL7U of UVSOR-II by use of a linearly polarized light with p-polarization. Many tiny specular cleaved regions as well as non-specular rough surface regions were coexisting on this fractured clean surface within the beam spot size of few hundred μm. Measurement was performed at hν between 17 and 7 eV by using the MBS A-1 hemispherical analyzer. The acceptance angle along the entrance slit of the analyzer was 14°.

The angle-resolved spectra (ARPES) of $SmB_6$ were measured at BL-9B of HiSOR at temperatures between 40 and 12 K by use of circularly polarized synchrotron radiation (SR) light at hν~35 eV with the hν resolution of 20 meV on a cleaved (100) surface. A single crystalline $SmB_6$ was fixed in a drilled hole of the sample holder by means of conductive epoxy (Muromacbond). By use of a pinpost cleaving of a single crystal sample with a dimension of 1x1x2 $mm^3$, a specularly cleaved region with a scale of ~0.5 mm was obtained at 82 K under the vacuum of ~$4x10^{-8}$ Pa in the analyzer chamber, in which the cleaved sample was further cooled down to 40~12 K under the vacuum of $8x10^{-9}$ Pa (at 12 K). The measurement was performed by use of SCIENTA R4000 hemispherical analyzer modified for spin polarized photoemission with the analyzer energy resolution set to 19 meV. The SR was incident onto the $SmB_6$ at ~50° from its surface normal.

For SP-ARPES, analyzer resolution was set to 40 meV and the acceptance angle along the slit was set to 1.5 degrees. The spin polarization was measured for photoelectrons emitted ~3° off the surface normal by use of an Fe-O VLEED spin detector [15,16] for the $\sigma^+$ and $\sigma^-$ circularly polarized light as well as horizontally and vertically polarized light as illustrated in Fig.1. The degree of the spin polarization was evaluated by changing the direction of the remanent magnetization of the Fe of this spin-detector by reversing the current through a coil as shown in Fig.1. The incidence angle of the emitted photoelectron from $SmB_6$ onto the Fe-O spin-detector was set to ~6.5° from the Fe-O surface-normal and the incident electron energy was decelerated down to ~6 eV. The diffracted electron intensity was measured by a channeltron. After SP-ARPES measurement, the sample was once retracted to the sample preparation chamber and exposed to a vacuum of $2x10^{-7}$ Pa for 20 minutes and again transferred back into the analyzer chamber where the angle-resolved spectra were measured again to check the surface contamination effect on the band dispersions.

## 3. Experimental results

### 3.1 Angle integrated spectra

First of all hν dependence of the results of $Sm_{0.85}Eu_{0.15}B_6$ is described. The behavior of the angle-integrated spectra with changing the temperature between ~200 and 10 K was rather similar between this material and $SmB_6$ except for the absolute $E_B$ of the peak structure (not shown). Figure 2 shows the hν dependence of the spectra near $E_F$ of $Sm_{0.85}Eu_{0.15}B_6$ measured at 15 K between hν=17 and 7 eV with the total energy resolution $\Delta E_K$ of 17 to 9 meV. Gradual change of the spectral shapes with hν is recognized. It is noticed that a prominent peak is observed for hν<10 eV at $E_B$~12 meV.

Figure 3 shows the spectrum of $Sm_{0.85}Eu_{0.15}B_6$ at hν=7 eV and 16 K just after the fracturing under $5x10^{-8}$ Pa at BL7U of UVSOR-II and the spectrum of the same surface measured 22 hours after the fracturing under continuous radiation from SR. Two spectra are tentatively normalized at the peak for a simple comparison purpose. If a background is subtracted in the region up to 80 meV from the spectra in Fig.3, the ~12 meV peak is noticeably enhanced relative to the intensity at $E_B$>20 meV with time.

3.2 Angle-resolved spectra

The $E_B$–$k_{\|}$ intensity plot of the angle resolved spectra (ARPES) measured at BL-9B of HiSOR at 35 eV and 30 K on a cleaved (001) surface is shown in Fig.4. Figures 4(a) and (b) show the ARPES and its second energy derivative, respectively, in the $E_B$ range from $E_F$ to 6 eV measured just few minutes after the cleavage. The white regions in the spectrum (a) and the dark regions in the second energy derivative (b) correspond to the region with high photoemission intensity. The abscissa $k_{\|}$ is the k value along the (100) direction. The normal emission at hν=35 eV corresponds to $k_z$ in the middle region between the Γ(0,0,0) and X(0,0,π/a) direction [9].

Clear doublet with almost negligible dispersion is observed near $E_F$ within $E_B$<200 meV. In addition, dispersing bands are observed within 2 eV from $E_F$ as recognized in Figs.4(a) and (b). Here a band with a bottom (or maximum $E_B$) at $E_B$~1.7 eV near ~0 $Å^{-1}$ is clearly resolved. Further, a slightly dispersing band is recognized around $E_B$~0.6 eV near ~0 $Å^{-1}$. An additional band is also recognized around $E_B$~0.6 eV in the vicinity of $k_{\parallel}$~−0.6 Å$^{-1}$. Besides a dispersing band is observed between 2.5 and 3.5 eV with a top (minimum $E_B$) near $k_{\parallel}$~ −0.6 Å$^{-1}$.

Figures 4(c) and (d) show the corresponding results measured half an hour later from the measurement (a) on the same sample surface. Here, the dispersing bands and structures between 0.5 and 2 eV are no more recognized in the angle resolved spectrum, whereas the dispersing band between 2.5 and 3.5 eV is still observed as before.

4. Spin-polarized and angle-resolved photoelectron spectra

The emitted photoelectrons from $SmB_6$ at 12 K were accepted within 1.5° along the slit and guided to the Fe-O VLEED spin detector [15,16] and SP-ARPES spectra were measured by reversing the magnetization direction of the Fe-O spin detector along the x direction shown in Fig.1. The spin polarization was measured parallel to the x direction for the circularly polarized light excitation in Figs.5(a) and (b). The results for linearly polarized excitation are shown in Figs.5(c) and (d). These single channel SP-ARPES measurements were performed at 3° off the surface normal direction in order to cover $k_{\parallel x}$ region between ~0.11 and ~0.18 Å$^{-1}$ to try to observe the possible spin polarization of the surface electron pocket around the $\bar{\Gamma}$ point near $E_F$ with the least broadening, though the surface states were reported to be rather weak and broad [10-12]. Although the pass energy of the analyzer was set to the same value of 5 eV, the resolutions of both photon monochromator and the electron energy analyzer were broadened twice to have higher counting rate. Then the energy resolution (FWHM) was set as ~50 meV for these SP-ARPES measurements.

Figures 5(a)(b)(c)(d) show the results for the $\sigma^+$ and $\sigma^-$ circularly polarized lights as well as for the vertical and horizontal linearly polarized lights. The measuring time to obtain each result ((a) to (d)) was between 30 minutes and 1 hour. Two peaks are clearly observed in the $E_B$ from $E_F$ to 0.2 eV. As observed for the $\sigma^+$ polarization excitation (a), the two peaks show opposite sign of spin polarization with negative sign for the peak near $E_B$=10–40 meV and positive sign for the second peak at $E_B$~150–160 meV. It is noticed that the magnitude of spin-polarization is around –0.40 for the first peak and +0.20 for the second peak. Since the spin polarization was estimated by $P_s=(I\uparrow-I\downarrow)/(I\uparrow+I\downarrow)$ in Figs.5(a)–(d), the contributions of the background from non 4f state may influence the experimentally evaluated $P_s$. If one can neglect spin polarization of the background, $P_s$ might be evaluated as –0.42 and 0.28, respectively after subtracting the background in Fig.5(a). When the helicity of the excitation light is switched to $\sigma^-$, the polarization of the first peak show a positive sign and the second peak shows a negative sign. Their magnitude is 0.50 and –0.25. After subtracting the background contribution, $P_s$ become 0.57 and –0.34, respectively. The difference of the absolute value of the polarization for the $\sigma^+$ and $\sigma^-$ may be mostly due to the possible difference of the circular polarization of the incident light, which is not accurately calibrated for this experiment. Then, the spin polarization $P_s$s will be represented by the averaged values as 0.50 and 0.31 with opposite signs for the two peaks. When the polarization of the excitation light was switched to linearly polarized light, no spin polarization was observed not only for the prominent two peaks but also in the smaller $E_B$ region down to $E_B$=0 eV, where the surface states around the $\bar{\Gamma}$ point is expected. After the full set of spin-polarized angle-resolved photoelectron measurement, the spin-integrated ARPES measurement was performed again, where the spectra were found to be not essentially changed in 4 hours before and after the SP-ARPES measurement.

5. Discussion

The hν dependence of the angle integrated spectra shown in Fig.2 is first discussed. It is noticed here that the prominent peak observed at 12 meV at hν=7 eV is weakened with increasing hν at the temperature of 15 K. The observed behavior of this peak near $E_F$ is most likely due to the hν dependence of the photoionization cross section (PICS) or matrix element effect as explained later. As already reported, the gap opens at $E_F$ below ~150 K [5]. As for the temperature dependence of the spectra at hν=7 eV (not shown here), the peak closest to $E_F$ in $Sm_{0.85}Eu_{0.15}B_6$ becomes sharper and its peak energy decreases gradually from 24 meV at 200 K to 12 meV at 15 K as if the spectral weight shifts to lower (smaller) $E_B$ with decreasing the temperature. In $SmB_6$, a similar temperature dependence is observed. Namely the peak shifts from 29 meV at 200 K to 18 meV at 15 K at hν=8.4 eV [5]. Very similar temperature dependence was clearly observed in the bulk sensitive Sm 4f spectral weight measured at hν~8 keV [Fig.1 of Ref.5]. The shift of the peak in the spectra at hν=7 and 8.4 eV is strongly correlated with the temperature-dependence of the Sm 4f states.

When hν decreases from 80 eV down to 7 eV, for example, the PICS of the Sm 4f state decreases dramatically with hν [17]. Although the PICS of the Sm 4f states is one order of magnitude larger than that of the Sm 5d states at 80 eV, it becomes comparable to or slightly lower than that of the Sm 5d states below hν~27 eV as judged from the PICSs of La 5d and Gd 5d states [17]. The Sm 4f PICS becomes much lower than those of Sm 5d and B 2sp states for hν<17 eV. The PICS of Sm 4f states is still larger than that of the Sm 5d states at hν=35 eV and more than three times larger than those of the B 2sp states above hν=40 eV. The peak observed in Fig.2 for hν<17 eV is therefore not dominated by the Sm 4f spectral weight but due to the hybridized states between the B 2sp, Sm 5d and Sm 4f states. The smaller value of $E_B$ of the peak in $Sm_{0.85}Eu_{0.15}B_6$ at low temperatures compared with that of $SmB_6$ at the same temperature and hν is in accordance with the smaller degree of Sm 4f–conduction band (B 2sp and Sm 5d) hybridization accompanied by the smaller gap opening compared with $SmB_6$ [18].

It is noticed in Fig.3 that the PES intensity of $Sm_{0.85}Eu_{0.15}B_6$ measured at hν=7 eV and 16 K is relatively decreased above $E_B$~25 meV after 22 hours in the vacuum of $5x10^{-8}$ Pa. Though not shown here, the PES spectral shape of this material at hν=17 eV shows very similar behavior at 14 K after 26 hours. Such results are hardly explicable by the PICS effect. More plausible interpretation is that the electronic structures with $E_B$>20 meV at low temperatures are largely suppressed with time with possible change of the surface quality, while the peak of the bulk origin is not so much suppressed. The comparison of the results in Figs.4 (a) and (c) confirms this interpretation as well. Namely surface derived states are gradually suppressed with time on the surface of fractured or cleaved $Sm_{0.85}Eu_{0.15}B_6$ and $SmB_6$.

In-gap surface states in the hybridization gap in the Kondo insulator $SmB_6$ at low temperatures are recently reported by high energy resolution ARPES [10-12] in the vicinity of $E_F$. The electric resistivity behavior at low temperatures was ascribed to the in-gap states crossing $E_F$. Negligible $k_z$ dependence suggested the surface origin of these in-gap states. In-gap metallic surface states with noticeable dispersions were reported around the surface $\bar{X}$ and $\bar{\Gamma}$ points [10-12]. Four rather large oval-shaped Fermi surfaces were observed around the $\bar{X}$ point [10]. The in-gap metallic surface states around the $\bar{\Gamma}$ point was much weaker in the first Brillouin zone (BZ) compared with those in the second BZ at hν=25 eV [10]. The two prominent peaks observed in Figs.4(a) and (c) at $E_B$~12 and ~150 meV are ascribable to the $^6H_{5/2}$ and $^6H_{7/2}$ final states. Although these states resulting from the Sm 4f states are clearly resolved in both Figs.4(a) and (c), the in-gap metallic surface states are not resolved even in Figs.4(b) and (d) possibly due to their low intensity on the surface presently investigated and the photon energy (hν=35 eV) selected in this experiment, though the wave number $k_{Fx}$ crossing $E_F$ was reported to stay almost constant irrespective of hν [10-12].

Although time dependence or surface condition dependence of the spectral shapes was not discussed in Refs.10-12, that was carefully checked in Ref.19 at

hν=21.2 eV, where time evolution of ARPES spectra along $\bar{M}-\bar{\Gamma}-\bar{M}$ at 6 K under $5x10^{-9}$ Pa was presented.

The non-metallic dispersing bands around the $\bar{\Gamma}$ point with their bottoms at $E_B$~1.7 and 2.3 eV in Fig.4(b) are also reported in this reference [19]. In our experiment, these dispersive bands observed in Fig.4(b) are no more observed in Fig.4(d) after half an hour from the measurement shown in Figs.4(a) and (b) and can be ascribed to surface states. Though polarity-driven origin was proposed for the B-2p metallic surface state derived from the $B_6$ terminated $SmB_6$ surface near the $\bar{\Gamma}$ point [12], we could not observed such metallic surface states on our sample surface possibly due to the difference in the surface quality or photon energy (hν=35 eV).

Now the SP-ARPES results are discussed. Since the PICS of the Sm 4f states is noticeably larger than those of the Sm 5d and B 2sp states at hν=35 eV, the two peaks observed in Fig.5 are understood to correspond mainly to the Sm 4f dominated $^6H_{5/2}$ (smaller $E_B$) and $^6H_{7/2}$ final states. Let $\ell$ and $\ell'$ indicate the azimuthal quantum number of the electron before and after the photoexcitation. According to the dipole excitation, it is known that $\ell'=\ell\pm1$. In the present case, $\ell$ is considered to be 3 and the photoelectron has either $\ell'=2$ or 4. It has been predicted for photoexcitation of Ce 4f electron by photons of 40.8 eV that photoexcitation probability to the $\ell'=4$ state is much larger (by more than several times) than that to the $\ell'=2$ state [20]. Therefore, we suppose that Sm 4f electrons are mostly excited as photoelectrons with $\ell'=4$.

Circularly polarized light with $\sigma^+$ ($\sigma^-$) polarization consists of photons with spin parallel (antiparallel) to the propagation vector of the light. Under the electric dipole approximation, the excitation operator of the $\sigma^+$ light propagating to the +z direction is proportional to x+iy, and therefore is proportional to $Y_1^1(\theta, \varphi)$ where $Y_\ell^m$ is the spherical harmonic function. A 4f electron whose magnetic quantum number is $\ell_z$ is excited by the light with circular polarization $\sigma^+$ to the photoelectron with the magnetic quantum number $\ell_z+1$. Based on our supposition that the photoelectron has $\ell'=4$, the dependence of the photoexcitation probability is $|\int Y^*{}_4^{(\ell_z+1)}\, Y_1^1\, Y_3^{\ell_z}\, d\Omega|^2$, which is proportional to $(\ell_z^2 + 9\,\ell_z + 20)/2$, for the $\sigma^+$ excitation. The ratio of this probability is calculated to be 28:21:15:10:6:3:1 for $\ell_z$ =3, 2, 1, 0, -1, -2, -3. This result indicates that 4f electrons with positive $\ell_z$ are preferentially photoexcited by the $\sigma^+$ circularly polarized light.

Then we assume, for simplicity, that the ground state of the $4f^6$ state is represented by $^7F_0$, whose total angular momentum is J=0. Then the total angular momentum of the final state should be the same as the angular momentum of the photoexcited hole j. When the final state is $^6H_{5/2}$, a 4f electron with j=5/2 is photoexcited. There are six j=5/2 states with magnetic quantum numbers of $j_z$=−5/2, −3/2, −1/2, 1/2, 3/2 and 5/2, with the wave function

$$|j=5/2,\ j_z\rangle = [(7-2j_z)/14]^{1/2} \cdot Y_3^{j_z-1/2}(\theta,\varphi)\uparrow - [(7+2j_z)/14]^{1/2} \cdot Y_3^{j_z+1/2}(\theta,\varphi)\downarrow \quad (1),$$

where ↑ and ↓ are the up and down spin functions, respectively. Then the wave function of the photoelectron becomes $Y_4^{j_z+1/2}(\theta,\varphi)\uparrow$ or $Y_4^{j_z+3/2}(\theta,\varphi)\downarrow$ for the $\sigma^+$ or $\sigma^-$ excitation with probability proportional to

$$\{(7-2j_z)/14\}\{(j_z-1/2)^2+9(j_z-1/2)+20\} \quad (2)$$

and

$$\{(7+2j_z)/14\}\{(j_z+1/2)^2+9(j_z+1/2)+20\} \quad (3),$$

respectively. The probability of finding the electron with wave function $Y_{\ell'}^{\ell_z'}(\theta, \varphi)$ in the present experimental geometry is

$|Y_{\ell'}^{\ell_{z'}}(50°, 0°)|^2$. (4)

Taking these factors into account, the spin polarization averaged over the six $j=5/2$ 4f states with $j_z=-5/2, \cdots, 5/2$ excited by the $\sigma^+$ light becomes –0.43. Therefore, the $^6H_{5/2}$ peak is expected to have spin polarization of –0.43 for fully polarized $\sigma^+$ light excitation. Similar calculation leads to spin polarization of the $^6H_{7/2}$ peak as +0.32.

Now let us more realistically consider that the ground state is a Kondo state consisting of both $f^5$ and $f^6$ states. There are two possible Kondo states, namely, $f^6+f^5L$ and $f^5+f^6\underline{L}$. Here L ($\underline{L}$) represents an electron (a hole) in the ligand electronic orbital. According to a preliminary calculation by using the Xtls code [21], the $f^6+f^5L$ initial state results in essentially the same spin polarization as calculated in the previous paragraph. On the other hand, the $f^5+f^6\underline{L}$ initial state results in a much smaller spin polarization. Therefore we consider that the present experimental results indicate that the Kondo state realized in this system is $f^6+f^5L$.

In this scheme, spin polarization is induced along any light incidence direction on the circularly polarized light excitation due to the dipole selection rules caused by the spin-orbit interaction and spin polarization is not expected for the excitation by the linear polarization light in agreement with the experimental results. This result demonstrated that the spin polarized Dirac cone state is not observed in the spectra observed in this experiment. In other words, completely linear polarization light excitation or fully unpolarized light excitation is inevitable to check the presence or absence of the Dirac cone state with intrinsic spin polarization in $SmB_6$ at low temperatures in the $E_B$ region within 20 meV from $E_F$, where an energy resolution better than 10 meV may be required for measurements. Since low detection efficiency of single channel spin detection cannot overcome the time–dependent surface quality change, higher efficiency spin detector will be inevitable in addition to realizing best quality ultrahigh vacuum condition to acquire complete set of information on two–dimensional spin polarization.

## 6. Conclusion and prospect

As demonstrated in the present experiment, the gradual change of the surface quality with time induces a serious change of the spectral shapes in the surface sensitive low hν photoemission at hν=35 eV in the $E_B$ region between $E_F$ and 2 eV, where most interesting surface states are expected to exist in this strongly correlated electron system, $SmB_6$. As experimentally revealed, the structures near $E_B$ of ~12–20 meV and ~150–180 meV survives even after the change of surface quality and are thought to be reflecting the bulk electronic states dominated by the 4f $^6H_{5/2}$ and $^6H_{7/2}$ components at hν=35 eV hybridized with the B 2sp and Sm 5d states. The whole results of the spin-polarized and angle-resolved spectra obtained here are consistently understood by considering the dipole selection rules for these states by the circularly and linearly polarized light excitation. The non-observation of the spin-polarization of the proposed Dirac cone state [6,7] in the present experiment may be either due to the change of the surface quality with time or due to the different origin of the surface metallic states. In order to study the spin polarization of the possible Dirac cone states in this surface sensitive material, short measuring time of SP-ARPES in two-dimensional k space under a very high quality ultrahigh vacuum as well as the use of fully linearly polarized or unpolarized light will be inevitable in addition to very high energy resolution better than 10 meV. If ellipsoidal light is employed, it may induce the spin polarization of the $^6H_{5/2}$ states obscuring the possible spin polarization of weak and tiny Dirac cone states.

Acknowledgement

The authors (S. S. and K. S.) are much obliged to Profs. M. Taniguchi and A. Kimura for their interest in this subject and general warm support for this experiment. S.S. also acknowledges Dr.C.Tusche and Prof.J.Kirschner for fruitful discussions and exchange of ideas on the future direction of spin-polarized angle resolved photoelectron spectroscopy on general solid state materials.

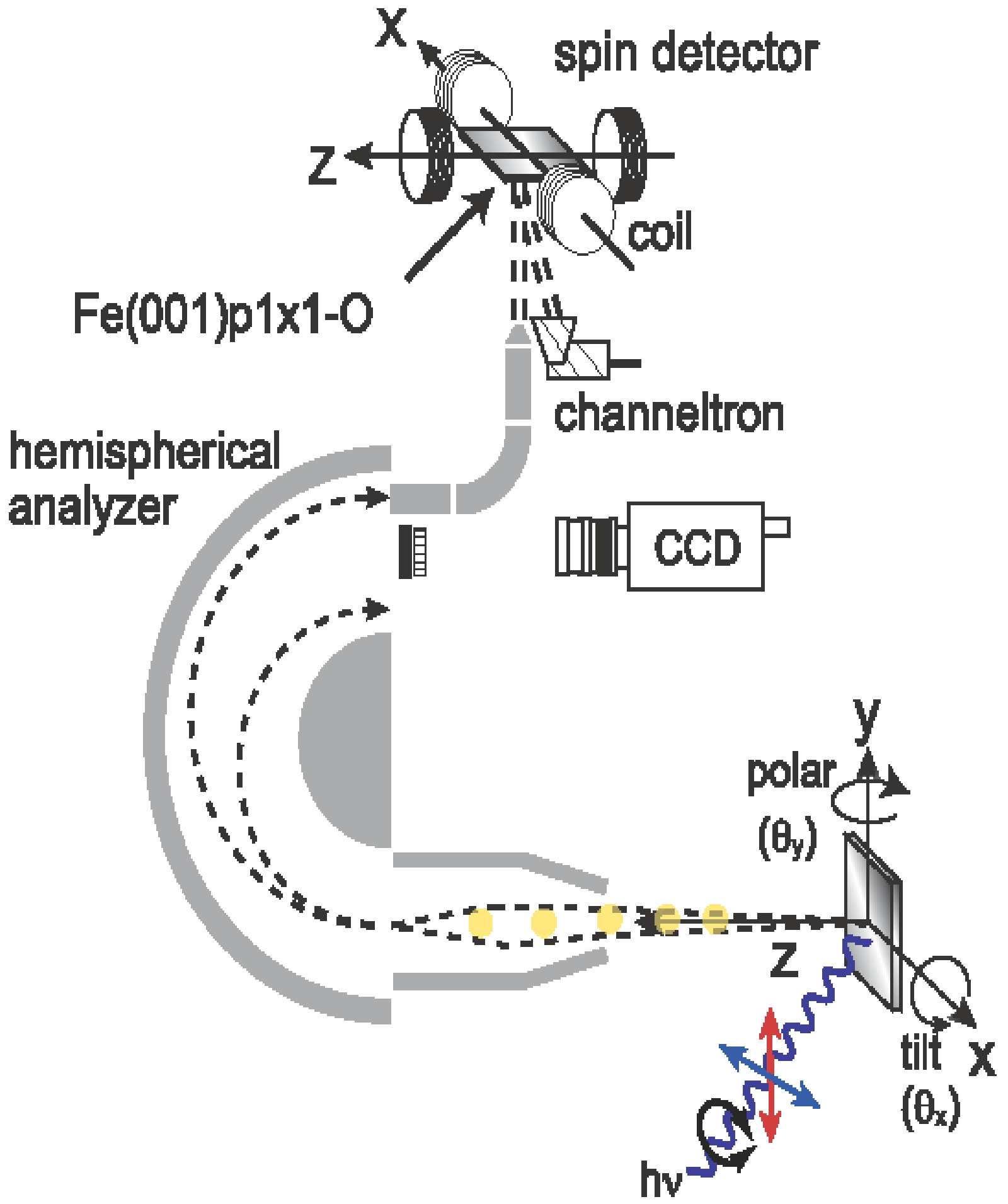

Fig.1 Experimental setup for spin-polarized angle-resolved photoelectron spectroscopy at BL-9B of HiSOR. The light was incident in the x–z plane and the incidence angle was set to ~50° from the z direction in the case of the normal photoemission.

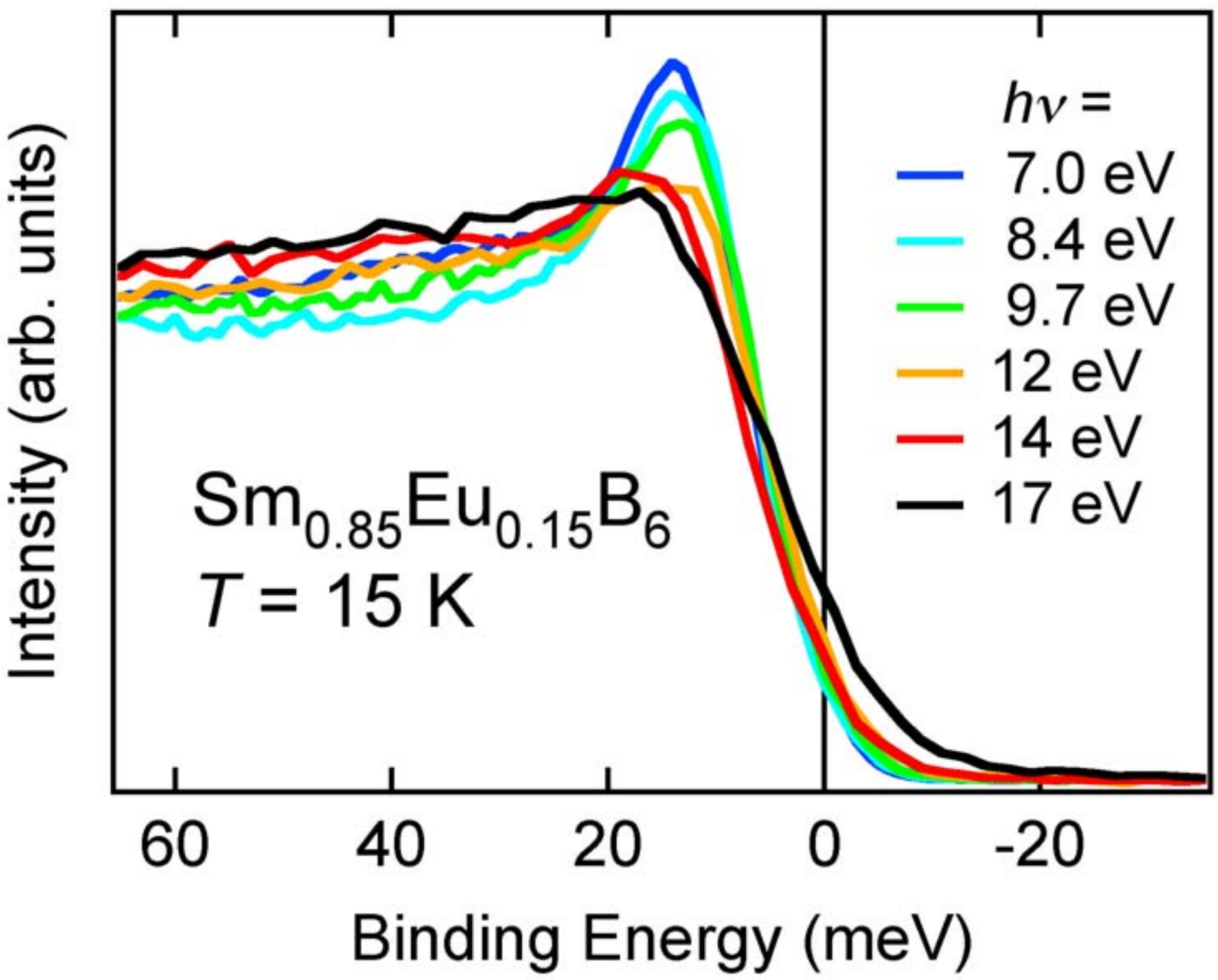

Fig.2 Angle integrated photoelectron spectra of $Sm_{0.85}Eu_{0.15}B_6$ measured at 15 K between hν=17 and 7 eV.

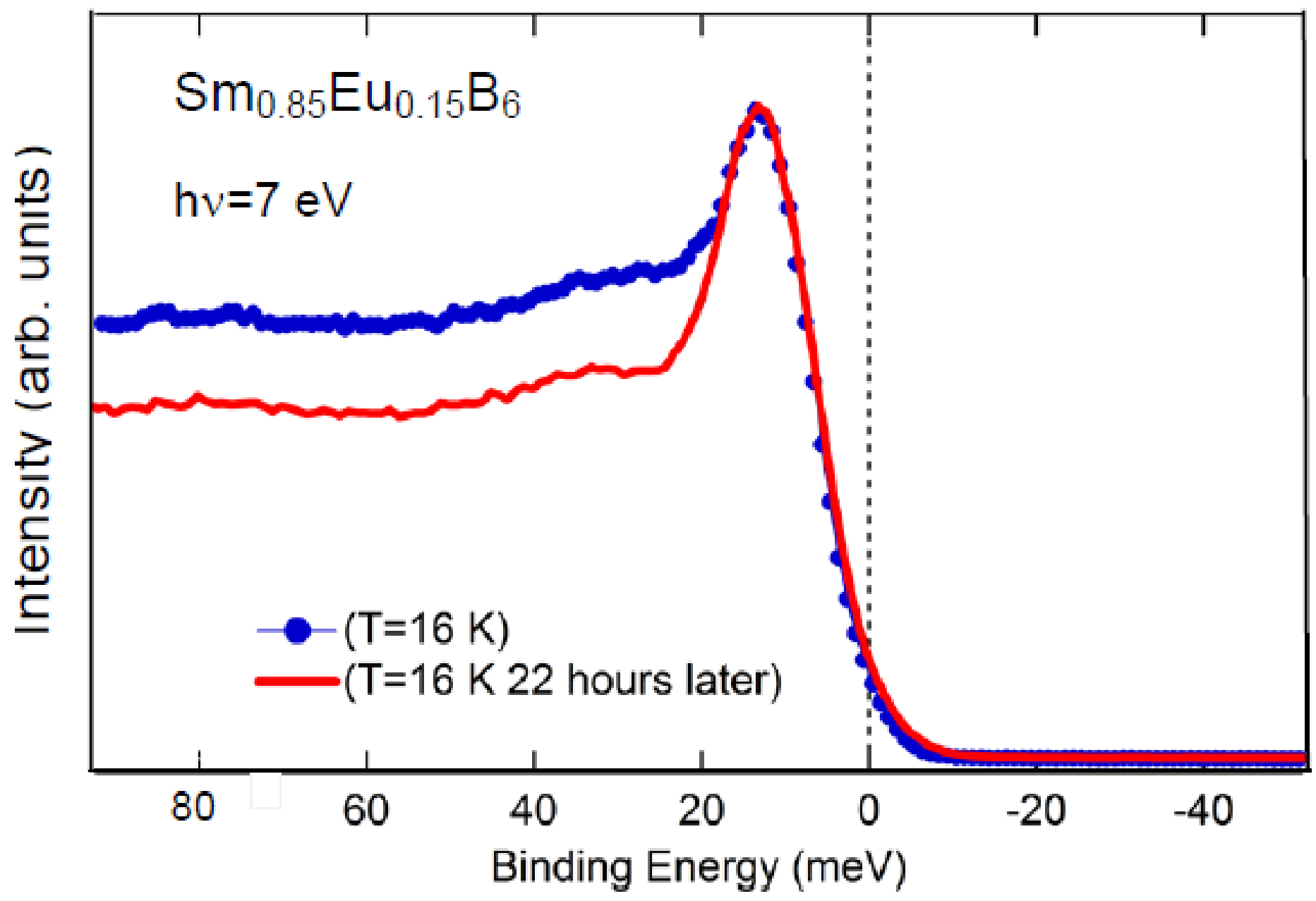

Fig.3 Spectral change of the angle integrated spectra of $Sm_{0.85}Eu_{0.15}B_6$ with time measured at hν=7 eV and 16 K under the vacuum of $5x10^{-8}$ Pa.

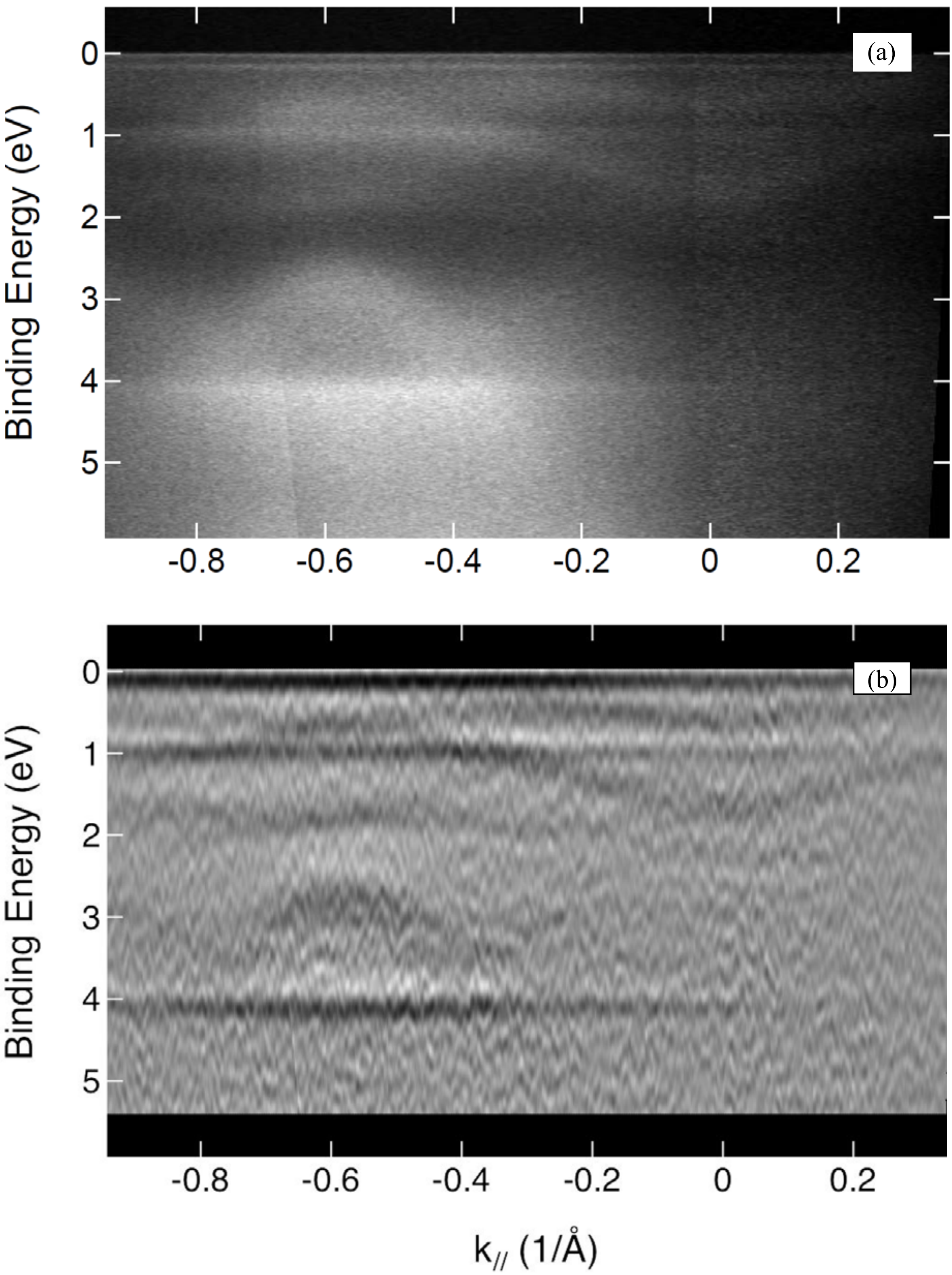
(a)
Binding Energy (eV)
0
1
2
3
4
5
-0.8
-0.6
-0.4
-0.2
0
0.2
(b)
Binding Energy (eV)
0
1
2
3
4
5
-0.8
-0.6
-0.4
-0.2
0
0.2
k// (1/Å)

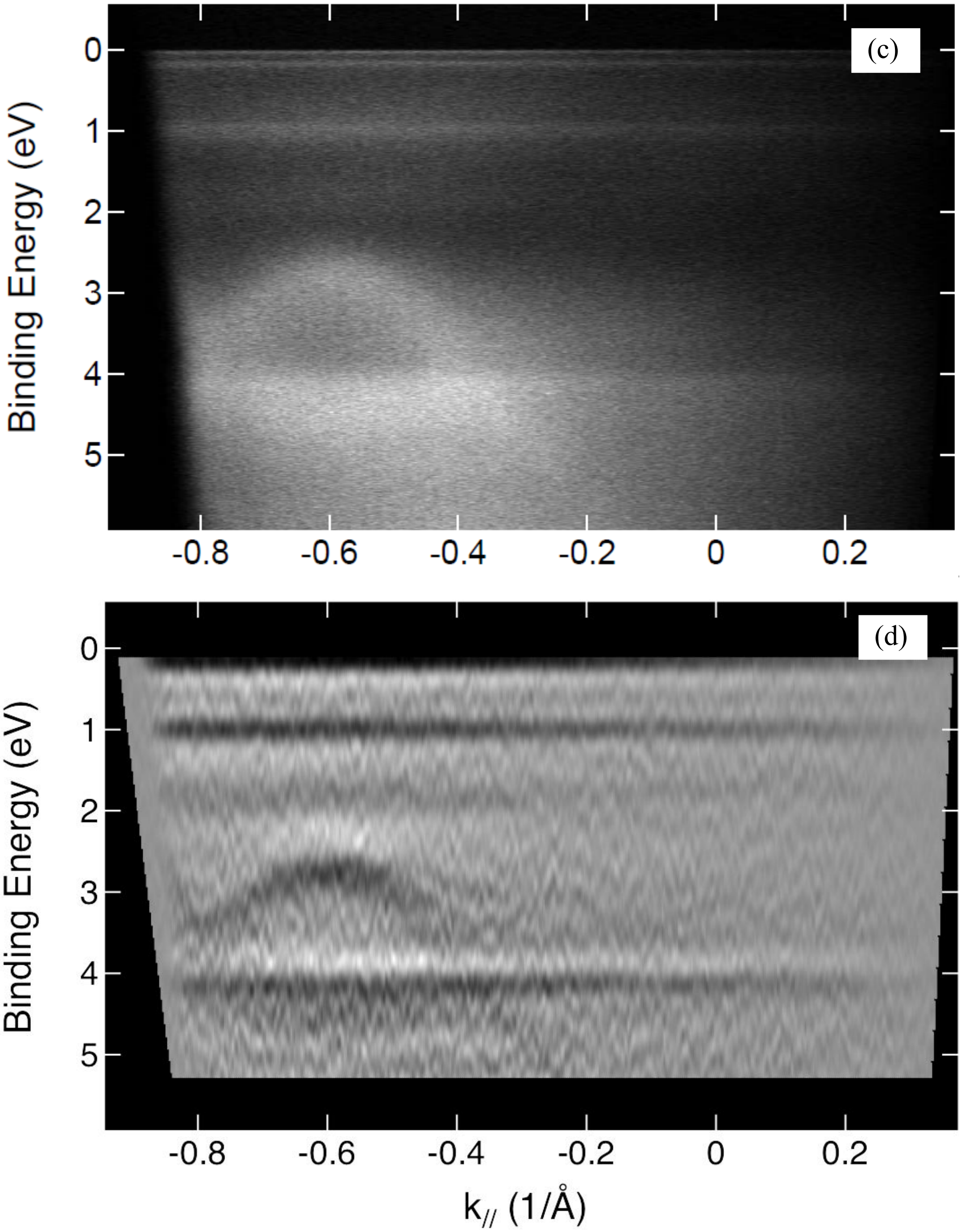

Fig.4 Angle-resolved photoemission spectra of $SmB_6$ measured at hν=35 eV at 12 K parallel to the (0, 0, 0)-(π/a, 0, 0) direction (or surface $\bar{\Gamma}-\bar{X}$ direction). (a) and (b) are raw spectrum and second energy derivative just after cleavage. (c) and (d) are those after 1 hour under the vacuum of ~$1\times10^{-8}$ Pa.

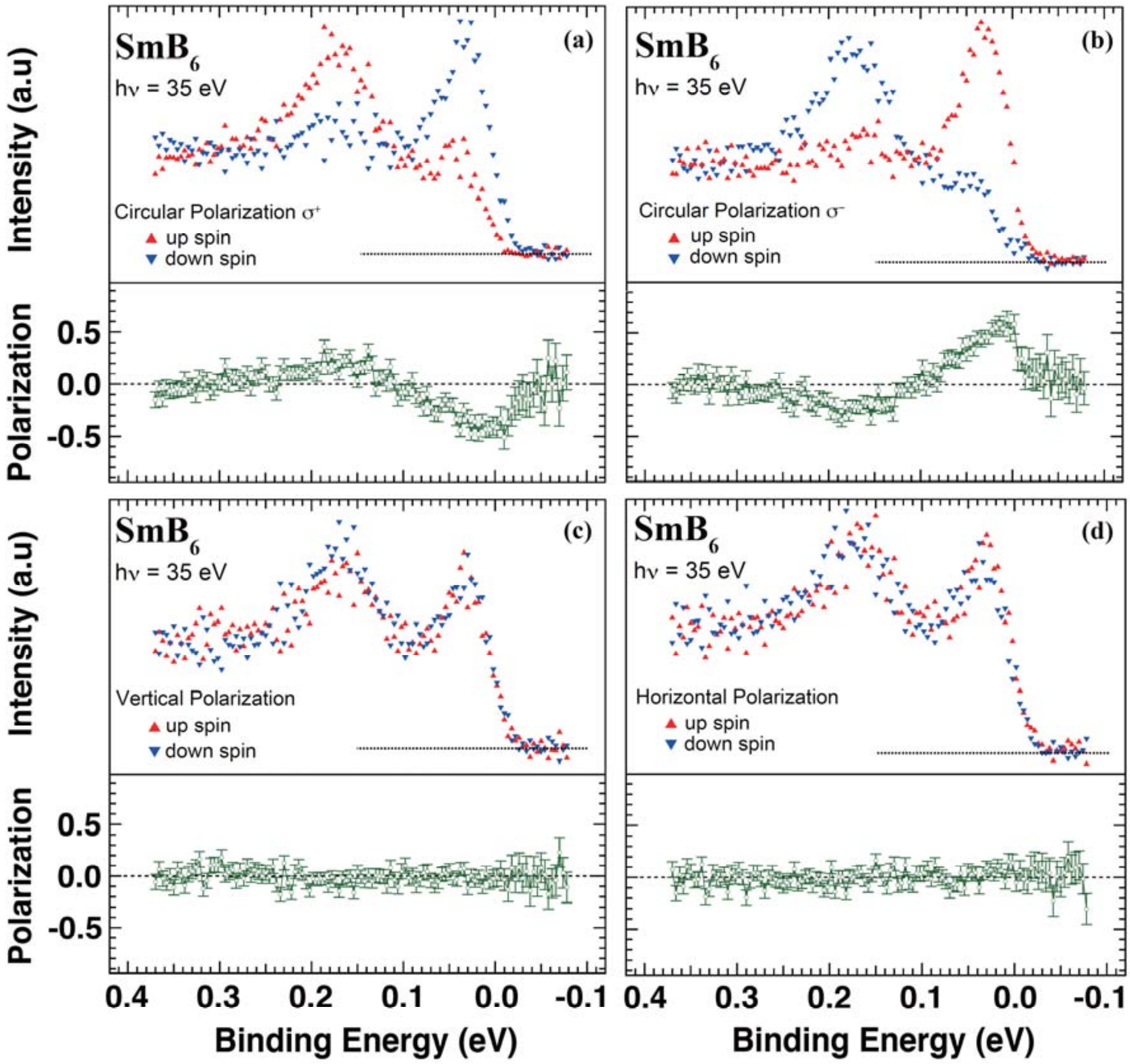

Fig.5 Spin-polarized and angle-resolved photoelectron spectra measured at 12 K and hν=35 eV after the measurement shown in Fig.4(b). $\sigma^+$, $\sigma^-$, linearly polarized light excitations for the spectra (a),(b),(c) and (d). The spin parallel to the x axis in Fig.1 was observed for the circularly polarized light excitation. Spin polarization was not observed for the linearly polarized light excitation.